\documentclass[fleqn,twoside]{article}
\usepackage[headings]{espcrc2}
\usepackage{amsmath,amssymb,alltt,axodraw}

\newcommand\ri{\mathrm{i}}
\newcommand\eg{e.g.\ }
\newcommand\ie{i.e.\ }
\newcommand\lbrac{\symbol{123}}
\newcommand\rbrac{\symbol{125}}

\newcommand\backsl{\symbol{92}}
\newcommand\bsg{$b\to s\gamma$}
\newcommand\ket[1]{\left| #1\right\rangle}
\newcommand\bra[1]{\left\langle #1\right|}
\newcommand\Code[1]{\ensuremath{\texttt{#1}}}
\newcommand\Var[1]{\ensuremath{\mathit{#1}}}

\title{Excursions into FeynArts and FormCalc}

\author{T.~Hahn\address{%
		Max-Planck-Institut f\"ur Physik,
		F\"ohringer Ring 6, D--80805 Munich, Germany},
	J.I.~Illana\address{%
		Departamento de F{\'\i}sica Te\'orica y del Cosmos, and
		Centro Andaluz de F{\'\i}sica de Part{\'\i}culas
		Elementales (CAFPE),
		Universidad de Granada,
		E--18071 Granada, Spain}}

\begin{document}

\begin{abstract}
Programming techniques which extend the capabilities of FeynArts and 
FormCalc are introduced and explained using examples from real 
applications.
\end{abstract}

\maketitle


\section{Introduction}

FeynArts \cite{FeynArts} and FormCalc \cite{FormCalc} are programs for
the generation and calculation of Feynman diagrams.  The main problem
they solve is the computation of the cross-section for a given process,
and indeed this is meanwhile a heavily automated procedure that requires
little user input.  Feynman diagrams have a lot of other uses, however,
and this note shows some examples of how to extend the capabilities of
FeynArts and FormCalc to produce results not within the `standard
canon.'

Although cash-strapped institutes may tend to believe otherwise, it is
actually a feature, not a bug, that FeynArts and FormCalc are largely
written in Mathematica, for it is precisely the availability of a
powerful language that makes it easy for the user to examine and modify
results, and also to extend the functionality of these systems.
Mathematica is particularly suited for the latter as it is technically
an expert system, where knowledge is added in the form of transformation
rules.  Thus one usually does not have to modify the existing program
code (possibly including re-compilation), but just add new rules.  
Sections \ref{sect:diagfilt}, \ref{sect:model}, \ref{sect:scripting}, 
and \ref{sect:notxs} show some real-life examples of such extensions.
Section \ref{sect:abbr} exhorts on the concept of abbreviations in 
FormCalc, where some important improvements have been made recently.


\section{Programming Diagram Filters}
\label{sect:diagfilt}

Question: What if FeynArts' diagram selection functions are not enough?
What if I want, say, only diagrams with a fermion loop?

Answer: Inspect the internal structure of the inserted topologies, \ie 
the \Code{InsertFields} output.  The outermost structure is a 
\Code{TopologyList}:
\begin{alltt}
TopologyList[__][\(t\sb1\), \(t\sb2\), ...]
\end{alltt}
Contained in this are topologies $t_i$ of the form
\begin{alltt}
Topology[_][__] ->
  Insertions[Generic][\(g\sb1\), \(g\sb2\), ...]
\end{alltt}
The generic insertions $g_i$ specify the fields running on each line of 
the topology, \eg
\begin{alltt}
Graph[__][Field[1] -> F, ...] -> \(c\sb{i}\)
\end{alltt}
Now recall that FeynArts distinguishes three levels of fields (generic, 
classes, particles) where the field's space-time properties are fixed at 
the lowest (generic) level \cite{methods}.

The fermion-ness of a field is a space-time property, so we need 
not go deeper than the generic level.  In order to look at generic-level 
diagrams, the selection function for fermion loops must first 
``propagate down'' into the generic level:
\begin{alltt}
FermionLoop[\,t:TopologyList[___][__]\,] :=
  FermionLoop/@ t\medskip
FermionLoop[\,(top:Topology[_][__]) ->
        ins:Insertions[Generic][__]\,] :=
  top -> TestLoops[top]/@ ins
\end{alltt}
The subsidiary function \Code{TestLoops} selects only the fields 
carried by propagators in the loop:
\begin{alltt}
TestLoops[top_] := TestFerm[\,Cases[top,
  Propagator[_Loop][_, _, field_] ->
    field]\,]
\end{alltt}
and hands on to \Code{TestFerm} which tests if, for a concrete 
diagram, the selected fields are indeed all fermions:
\begin{alltt}
TestFerm[fields_][gi_ -> ci_] :=
  (gi -> ci) /;
    MatchQ[fields /. List@@ gi, \lbrac{}F..\rbrac{}]\medskip
TestFerm[_][_] := Sequence[]
\end{alltt}
Note this typical (for Mathematica) construct: the first instance 
handles diagrams positively identified as having a fermion loop,
and the second one is the `fall-through' for all other cases.


\section{Tweaking Model Files}
\label{sect:model}

Question: Is there a good way to make (small) changes to an existing
model file?

Answer: It is bad practice to copy the model file and modify the copy,
for two reasons:
\begin{enumerate}
\item
It is typically not very transparent what has changed.

\item
If the original model file changes (\eg due to bug fixes), these
do not automatically propagate into the derivative model file.
\end{enumerate}
Better: Create a new model file which reads the old one and modifies the
particles and coupling tables.  To this end one needs to know that the 
model file defines two main objects:
\begin{itemize}
\item
\Code{M\$ClassesDescription} is the list of particle definitions,

\item
\Code{M\$CouplingMatrices} is the list of couplings
of the form \Code{C[\Var{fields}] == \Var{expr}}.
\end{itemize}

As an example consider introducing enhancement factors for the 
$b$--$\bar b$--$h_0$ and $b$--$\bar b$--$H_0$ Yukawa couplings in the 
MSSM.  The new model file is both compact and makes it immediately
obvious what has changed:
\begin{alltt}
Block[\lbrac{}$Path = $ModelPath\rbrac{}, << MSSM.mod]\medskip
EnhCoup[\,(lhs:C[F[4,\lbrac{}g_,_\rbrac{}], -F[4,_],
                 S[h:1|2]]) == rhs_\,] :=
  lhs == Hff[h,g] rhs\smallskip
EnhCoup[other_] = other\medskip
M$CouplingMatrices =
  EnhCoup/@ M$CouplingMatrices
\end{alltt}
Note that the enhancement factors depend on the fermion generation.  
This is because the couplings are defined at the classes level, \ie for 
the class \Code{F[4]} of down-type quarks $\{d,s,b\}$, not for the 
bottom quark alone.  Thus one needs to set \Code{Hff[h,1]} = 
\Code{Hff[h,2]} = 1.

A printout of the new Feynman rules can be obtained with the
\Code{WriteTeXFile.m} program that comes with FeynArts.


\section{Scripting Mathematica}
\label{sect:scripting}

Question: How can I do efficient batch processing with Mathematica?

Answer: Put everything into a script, using sh's Here documents:
\begin{alltt}
#! /bin/sh ......... {\it Shell Magic}
math << \backsl{}_EOF_ ..... {\it start Here document}
  << FeynArts`
  << FormCalc`
  top = CreateTopologies[...];
  ...
_EOF_ .............. {\it end Here document}
\end{alltt}
Everything between ``\verb=<< \=\Var{tag}'' and ``\Var{tag}''
goes to Mathematica as if it were typed from the keyboard.  Note the 
``\verb=\='' before \Var{tag}, it makes the shell pass everything 
literally to Mathematica, without shell substitutions.  This is
important because Mathematica uses many characters which have a 
special meaning to the shell, such as \Code{\$}, \Code{[},
\Code{\lbrac}, etc.

The advantages of this method are:
\begin{enumerate}
\item
Everything is contained in one compact shell script (i.e., a text file), 
even if it involves several Mathematica sessions.

\item
Such a script can easily be run in the background, or combined with 
utilities such as make.

\item
One can seamlessly combine Mathematica and shell programming.  A 
slightly subtle issue is how to get shell variables such as command-line
arguments into Mathematica even though substitutions in the Here file
are turned off through the use of \Code{\backsl} in
\Code{\backsl}\Var{tag}.  Solution: pass them on the command-line when
invoking Mathematica:
\begin{verbatim}
#! /bin/sh
math -run "arg1=$1" ... << \END
  ...
END
\end{verbatim}
\end{enumerate}
Debugging hint: the \Code{-x} flag makes sh echo every statement,
it can be added after the Shell Magic:
\begin{verbatim}
#! /bin/sh -x
\end{verbatim}


\section{Not the Cross-Section}
\label{sect:notxs}

Question: Can I get things out of FormCalc other than the cross-section?  
Can I, for example, compute the Wilson coefficients for \bsg?

Answer: The relevant operators for \bsg, the prefactors of which are 
the Wilson coefficients, are
\begin{gather}
O_7 = \frac{e}{16\pi^2} m_b 
  \bra{s_j}\omega_+\sigma_{\mu\nu}\ket{b_i}
  \delta_{ij} F^{\mu\nu} \\
O_8 = \frac{g_s}{16\pi^2} m_b
  \bra{s_j}\omega_+\sigma_{\mu\nu}\ket{b_i} 
  T^a_{ij} G_a^{\mu\nu}
\end{gather}
Generating the partonic diagrams and amplitudes is of course a standard 
exercise:
\begin{alltt}
tops = CreateTopologies[1, 1 -> 2]\smallskip
ins = InsertFields[\,tops,
  F[4,\lbrac{}3\rbrac{}] -> \lbrac{}F[4,\lbrac{}2\rbrac{}],\,V[1]\rbrac{}\,]\smallskip
amp = CalcFeynAmp[\,CreateFeynAmp[ins],
  FermionChains -> Chiral\,]
\end{alltt}
For the $O_8$ coefficient, the photon \Code{V[1]} has to be replaced by 
the gluon \Code{V[5]}.  The \Code{FermionChains} option ensures that the 
result contains the chiral Dirac chains needed to read off the Wilson 
coefficients, rather than the default Weyl chains.

\Code{CalcFeynAmp} collects Dirac chains and colour matrices in a
function \Code{Mat}.  For identifying the operators it is thus simplest
to construct a function \Code{id} that replaces \Code{Mat}.

The identification of the operators is slightly nontrivial because
FormCalc strives to reduce the number of Lorentz indices as much as
possible and thus turns $\omega_\pm\gamma^\mu\gamma^\nu$ into
$\omega_\pm\gamma^\mu$.  This can be reversed by applying the Gordon 
identity
\begin{multline}
\bra{s_2}\omega_\pm\gamma^\mu\ket{s_1} =
\frac{(p_1 + p_2)^\mu}{m_1} \bra{s_2}\omega_\pm\ket{s_1} + \\
\frac{\ri (p_2-p_1)_\nu}{m_1}
  \bra{s_2}\omega_\pm\sigma^{\mu\nu}\ket{s_1}\,.
\end{multline}
Since this is contracted with $\varepsilon_\mu$, the photon/gluon 
polarization vector, we can use the momentum-space correspondence
$\sigma^{\mu\nu} F_{\mu\nu}\to 2 \sigma^{\mu\nu} (p_2-p_1)_\nu 
\varepsilon_\mu$ to identify the $F_{\mu\nu}$-term, thus
\begin{multline}
\bra{s_2}\omega_\pm\rlap{/}\varepsilon\ket{s_1} =
\frac 2{m_1}\,\varepsilon\cdot p_1 \bra{s_2}\omega_\pm\ket{s_1} + \\
\frac{\ri}{2 m_1} \bra{s_2}\omega_\pm\sigma^{\mu\nu}\ket{s_1}
  F_{\mu\nu}\,.
\end{multline}
\begin{alltt}
id[r_. DiracChain[s2_Spinor, om_, eps_,
           s1:Spinor[p1_, m1_, _]]] :=
  2/m1 r Pair[eps, p1] *
    DiracChain[s2, om, s1] +
  I/(2 m1) id[r sig[om]]
\end{alltt}
It is not even necessary to explicitly write out the 
$\sigma^{\mu\nu}$-term since its only purpose is to match the operators:
\begin{alltt}
id[r_. sig[om_] SUNT[i_, j_]] :=
  r O7[om]/(EL MB/(16 Pi^2)) \smallskip
id[r_. sig[om_] SUNT[a_, i_, j_]] :=
  r O8[om]/(GS MB/(16 Pi^2))
\end{alltt}
Now we can apply this function to the amplitude:
\begin{alltt}
amp = Plus@@ amp //. Abbr[] /. Mat -> id\smallskip
c7 = Coefficient[amp, O7[6]]
c8 = Coefficient[amp, O8[6]]
\end{alltt}

Using FormCalc's output functions it is also pretty straightforward to 
turn these expressions into Fortran code:
\begin{alltt}
file = OpenFortran["bsgamma.F"]\smallskip
WriteString[file,
  SubroutineDecl["bsgamma(C7, C8)"] <>
  "\backsl{}tdouble complex C7, C8\backsl{}n" <>
  "#include \backsl{}"model.h\backsl{}"\backsl{}n" <>
  "#include \backsl{}"looptools.h\backsl{}"\backsl{}n"]\smallskip
WriteExpr[file, \lbrac{}C7 -> c7, C8 -> c8\rbrac{}]\smallskip
WriteString[file, "\backsl{}tend\backsl{}n"]\smallskip
Close[file]
\end{alltt}


\section{Abbreviations}
\label{sect:abbr}

The automated introduction of abbreviations is one of the key concepts
in FormCalc.  It is crucial in rendering an amplitude as compact as
possible.  The main effect comes from three layers of recursively
defined abbreviations, introduced when the amplitude is read back from
FORM, \ie during \Code{CalcFeynAmp}.  For example:
\begin{center}
\begin{picture}(200,60)(0,12)
\SetScale{.8}

\Text(0,60)[bl]{\Code{AbbSum29 = Abb2 + Abb22 + Abb23 + Abb3}}
\SetOffset(108,12)
\EBox(-19,55)(19,72)
\Line(-19,55)(-85,41)
\Line(-85,41)(-85,28)
\Line(19,55)(85,41)
\Line(85,41)(85,28)

\Text(0,24)[b]{\Code{Abb22 = Pair1 Pair3 Pair6}}
\SetOffset(129,12)
\EBox(-19,25)(19,42)
\Line(-19,25)(-79,13)
\Line(-79,13)(-79,0)
\Line(19,25)(79,13)
\Line(79,13)(79,0)

\Text(0,0)[b]{\Code{Pair3 = Pair[e[3],\,k[1]]}}
\end{picture}
\end{center}
Written out, this abbreviation is equivalent to
\begin{footnotesize}
\begin{alltt}
Pair[e[1],\,e[2]] Pair[e[3],\,k[1]] Pair[e[4],\,k[1]] +
Pair[e[1],\,e[2]] Pair[e[3],\,k[2]] Pair[e[4],\,k[1]] +
Pair[e[1],\,e[2]] Pair[e[3],\,k[1]] Pair[e[4],\,k[2]] +
Pair[e[1],\,e[2]] Pair[e[3],\,k[2]] Pair[e[4],\,k[2]]
\end{alltt}
\end{footnotesize}

In addition to these abbreviations assigned by \Code{CalcFeynAmp},
FormCalc introduces another set of abbreviations for the loop integrals
when generating Fortran code, \ie during \Code{WriteSquaredME}.


\subsection{Categories}

Both of the aforementioned types of abbreviations, but in particular the
latter, are costly in CPU time.  It is thus key to a decent performance
that the abbreviations are grouped into different categories:
\begin{enumerate}
\item Abbreviations that depend on the helicities.
\item Abbreviations that depend on angular variables.
\item Abbreviations that depend only on $\sqrt s$.
\end{enumerate}
Correct execution of the different categories guarantees that almost no
redundant evaluations are made.  For example, for a $2\to 2$ process
with external unpolarized fermions, statements in the innermost loop
over the helicities are executed $2^4$ times as often as those in the
loop over the angle.  This technique of moving invariant expressions out
of the loop is known as `hoisting' in computer science.


\subsection{Common Subexpressions}

Another optimization method, common subexpression elimination, is
implemented in the function \Code{OptimizeAbbr} and can yield some
additional 10--30\% speed-up.  It works in two steps.  First, redundant
parts are removed, \eg the abbreviations
\begin{verbatim}
AbbSum2 -> Abb9 - Abb87
AbbSum8 -> Abb9 - Abb13 - Abb74 - Abb87
AbbSum9 -> Abb9 + Abb13 + Abb74 - Abb87
\end{verbatim}
are replaced by
\begin{verbatim}
AbbSum2 -> Abb9 - Abb87
AbbSum8 -> AbbSum2 - Abb13 - Abb74
AbbSum9 -> AbbSum2 + Abb13 + Abb74
\end{verbatim}
In the second step, common parts are put into temporary variables,
thereby simplifying the last lines further to
\begin{verbatim}
AbbSum2 -> Abb9 - Abb87
help1 -> Abb13 + Abb74
AbbSum8 -> AbbSum2 - help1
AbbSum9 -> AbbSum2 + help1
\end{verbatim}
Both optimization techniques together make the generated code 
essentially as fast as hand-tuned code.


\subsection{The Abbreviate Function}

The new \Code{Abbreviate} function extends the advantages of the 
abbreviation system to arbitrary expressions.  Its usage is for example:
\begin{verbatim}
abbrexpr = Abbreviate[expr, 5]
\end{verbatim}
The second argument, 5, determines the level below which abbreviations
are introduced.  The level determines how much of expression is
`abbreviated away,' \ie how much of the structure is preserved.  In the
extreme, for a level of 1, the result is just a single symbol. 
Abbreviationing also has the `side effect' that duplicate expressions 
are replaced by the same symbol.

This new type of abbreviations for subexpressions has to be retrieved
separately from the other ones with \Code{Subexpr[]}.

The most important option of \Code{Abbreviate} is \Code{Preprocess}.
It is used \eg as follows:
\begin{alltt}
abbrexpr = Abbreviate[\,expr, 5,
  Preprocess -> Simplify\,]
\end{alltt}
and specifies a function which is applied to each subexpression before 
the abbreviations are introduced.

At some 30 sec.\ execution time for \Code{Abbreviate}, the typical
speed-up was a factor 3 in MSSM calculations.


\subsection{Auxiliary functions}

When simplifying large expressions, it is often desirable to have a
replacement for \Code{Simplify} which is faster by performing only
specific simplifications.  This is because \Code{Simplify} is quite 
efficient on short expressions but increasingly slow on longer ones.

\Code{OnSize} constructs a special function for simplification that 
does different things depending on the size of its argument.  For 
example,
\begin{verbatim}
f = OnSize[100, Simplify,
           500, DenCollect,
           Pool]
\end{verbatim}
If the \Code{LeafCount} of the argument of \Code{f} is \\
$\bullet$ below 100, \Code{Simplify} is used, \\
$\bullet$ between 100 and 500, \Code{DenCollect} is used, \\
$\bullet$ above 500, \Code{Pool} is used.

\Code{DenCollect} collects terms with denominators that are identical 
up to a numerical constant.

\Code{Pool} combines terms with common factors.  Unlike 
\Code{Factor}, it looks at the terms pairwise and can thus do
$a b + a c + d \to a (b + c) + d$ fast.  \Code{Pool} will not factor out 
very small expressions because the effect of this on the size of the 
whole expression is typically not worth the effort.


\section{Summary}

We presented several examples of programming techniques which enable the
user to extend the capabilities of FeynArts and FormCalc with 
comparatively little effort:
\begin{itemize}
\item programming diagram filters,
\item applying (small) changes to model files,
\item scripting Mathematica,
\item computing quantities other than the cross-section.
\end{itemize}
In addition, we elucidated the important concept of abbreviations in 
FormCalc which is extended to arbitrary expressions by the new 
\Code{Abbreviate} function.


\section*{Acknowledgements}

TH thanks Granada University for kind hospitality during part of this 
work.


\end{document}